# Exploring the Pathways of Adaptation an Avatar 3D Animation Procedures and Virtual Reality Arenas in Research of Human Courtship Behaviour and Sexual Reactivity in Psychological Research

*Jakub Binter, Kateřina Klapilová, Tereza Zikánová, Tommy Nilsson, Klára Bártová, Lucie Krejčová, Renata Androvičová, Jitka Lindová, Denisa Průšová, Timothy Wells, and Daniel Riha*

**Abstract**

There are many reasons for utilising 3D animation and virtual reality in sexuality research. Apart from providing a mean with which to (re)experience certain situations there are four main advantages: a) bespoke animated stimuli can be created and customized, which is especially important when researching paraphilia and sexual preferences; b) stimulus production is less expensive and easier to produce compared to real world stimuli; c) virtual reality allows us to capture data such as physiological reasons to stimuli, that we would not be able to otherwise (without resorting to self-report measures which are especially problematic in this research domain); d) ethical, legal, and health and safety issues are less complex since neither physical nor psychological harm is caused to animated characters allowing for the safe presentation of stimuli involving vulnerable targets. The animation sub-group has been exploring so far several production quality levels and various animation procedures in a number of available software. The aim is to develop static as well as dynamic, interactive sexual stimuli for sexual diagnostic and therapeutic purposes. We are aware of number of ethical issues related to the use of virtual reality in proposed research are analysed in this chapter.

*****

The processing power of computers has been growing exponentially, propelling the emergence of a whole new set of tools for displaying and experiencing digital data. Text-based interfaces were gradually replaced by moving images before computers finally grew able to handle large sets of 3D coordinates, enabling the first interactive experiences taking place in virtual space.

More recently, advances in technology that harness the increasing power of computer hardware has made great strides toward making virtual reality (VR) environments feel as believable as possible.

Although VR has traditionally been considered an enabling technology for videogames, given its versatility and increasing sophistication, it comes perhaps as no surprise that VR methodology is quickly gaining a foothold in areas outside entertainment.

Military, education and healthcare are but a few examples of fields where VR technology experiences a significant growth. The ability to visualize and re-enact practically any real world scenario from within a lab or a training facility has proven to be exceedingly useful whilst being safer, less costly and far more flexible than more traditional experimental or training methods. Moreover research areas such as psychiatry are frequently utilizing VR as a diagnostic tool allowing scientists to produce accurate diagnosis by utilizing it in combination with other methods such as MRI scans, psychophysiological measures, self-reported data, observational inputs (by third person).[1]

In 2015, National Institute of Mental Health has been established as a new centre of research and leading institution in Czech Republic for psychiatric treatment as well as psychological and diagnostic research. The institute comprises more than thirty research groups with various specialisation. We are an inter-disciplinary research group of Evolutionary Sexology and Psychopathology, that is interested in several research topics ranging from investigating factors relating to sexual orientations, sexual activity preferences (e.g., sado-masochism), and sexual pathology, and behavioural aspects of sexual relations such as non-verbal behaviour, power and dominance. Currently, we are working to develop techniques in 3D animation and virtual reality that can be applied in our studies. Our projects demand us to recreate various social interactions that cannot be manifested in reality, in VR, to later serve as stimuli in an experimental study. This goal introduced us to a number of technical challenges stemming from the fact that we needed to accurately recreate human (non-verbal) behaviour, including gesticulations and facial expressions, with a high degree of reality.

Because manual creation of such animations requires a considerable time investment and the qualitative outcome of the process can vary, therefore we currently explore the idea of developing stimuli using motion sensors which allow us to record the given social interactions performed by real actors and then translate the recordings into animations. We believe that systems such as motion sensing device[2] in combination with relevant motion capture software packages, e.g. Faceshift or iPiSoft, may represent a viable framework allowing us to record various social interactions for their later accurate recreation in virtual reality. It is also considerably less expensive and more time efficient. Once generated, animations can be mapped over any 3D figure representing a virtual human, thus allowing for the speedy generation of different versions of any potential stimuli.

The potential benefits of recording human animations using motion sensors are further amplified when considering that such recordings can be translated in more robust 3D packages, such as Maxon Cinema 4D or Autodesk MotionBuilder, for further adjustments since these provide the user with ready-to-use motions and possibilities of blending existing ones with those developed by using motion capture.

Generating realistic virtual human characters is another significant challenge for our research. However the time generally required to create and manually texture sufficiently detailed 3D models prompted us to investigate alternative approaches. Solutions such as DAZ Studio or MakeHuman were found to offer a comparatively time efficient framework allowing us to procedurally transform basic humanoid models into desirable characters with relative ease.

Virtual characters can be brought to life through animations captured with a motion sensor and can finally be used to populate interactive virtual environments using the Unity 3D game engine. Our evaluation and estimation - comparison of costs, time efficiency, easy and intuitive operating, possibility to transfer data to other software, error rate, and hardware setting - suggests that such a pipeline might provide us with the most viable results.

Bellow we present an example of two research designs (dealing with timely topics in the field of sexology), where the possible application was discussed for years but the technology was not in reach and/or the software was not on today´s level of accuracy and graphics to serve well to the research aims.

## 2. Current Applications

### 2.1. Application 1: The Study of Ontogeny of Nonverbal Courtship Behaviour in Women

One of the most important choices made by an individual is the selection of a sexual partner.[3] Previous studies have described a system of non-verbal communication labelled ‘courtship’ that facilitates successful access to individuals of the opposite sex.[4] A number of studies[5] were focused on courtship behaviour especially in adults but there is minimal knowledge about development of non-verbal behaviour in childhood and young adulthood. The research is focused on ontogeny of female courtship behaviour during the lifespan which is a topic that has not been investigated in detail yet, but has important implications for treatment of various paraphilic disorders. Particularly, it is unknown in what age category the courtship behaviour appears, how it is modified during the lifespan and if there are inhibitory mechanisms that allow men to detect and sexually react exclusively on nonverbal courtship displays of women in reproductive age. Therefore, in this research, we focus on a) detection and animation of pure non-verbal behavioural displays in females of various age categories, and b) on perception of courtship in different age groups by male raters.

Therefore we decided to use motion capture followed by 3D animation to capture the nonverbal behaviour of six age categories of girls and women (age categories chosen to cover important periods of psychosocial development in females[6]). The intention is to combine movements obtained from motion capturing sensors with avatars of physical features corresponding to various age (in experimental setting) to test the reactions on pure nonverbal irrespective of variability in physical traits of target stimuli.

In the first step, the actors-volunteers representing each age category will be asked to present non-verbal behaviour displays of courtship during which movements are tracked by motion sensing sensors. The instruction are as follows: 'Imagine an attractive man and act in a way you would gain the attention of a potential "partner"'. The sub-adult volunteers[7] will be asked to interact with an unknown but sympathetic adult person of the opposite sex in a friendly manner.[8]

Then, we intend to create the stimuli rated by men of more age groups under the experimental conditions as follows:

1. Sub-adult avatar animated with courtship non-verbal displays of a female of the same age category,

2. Sub-adult avatar animated with courtship non-verbal displays of an adult female,

3. Adult-looking avatar animated with courtship non-verbal displays of sub-adult female,

4. Adult-looking avatar animated courtship non-verbal displays of an adult female.

If adult males rate adult female avatars animated with courtship non-verbal displays of sub-adult females as comparably attractive to adult avatars animated with non-verbal displays of adult female we would reveal that courtship is learned through imitation of post-pubertal models (parents, etc.). If the adult avatar animated with behaviour of a sub-adult female is rated as less attractive than the adult avatar animated with adult behaviour, it supports the theory favouring the explanation that courtship develops from age specific general social behaviour.

### 2.2. Application 2: The Study of Sexual Variation: Prevalence across Population, Physiological and Neural Correlates During Experimental Exposure to Erotic Stimuli

In the next project, we plan on using the VR for development of stimuli that would allow us to determine whether an individual has paraphilic sexual preferences - with a specific focus on Paedophilia. Particularly, we test the theory that the etiology of paedophilia is linked to inappropriate content of sexual motivational system.[9] Namely, there should be the appetence of non-verbal behavioural displays presented by children in affiliative situation that are false detected/misinterpreted as courtship behaviour and have a potential to start-up sexual reaction, whilst in normal population (heterosexual males attracted to adult women in reproductive age) these displays inhibit sexual reaction as well as negative proceptivity - rejective or startled reaction. In the project, the target subjects will be presented in VR by animated avatars with appropriate nonverbal displays[10] while the psychophysiological reaction of male subjects is measured (PPG,[11] GSR,[12] ECG[13]).

To create the behavioural displays patterns of stimuli, we will ask female actors to act out positively - flirtatiously (friendly for children) towards a potential sexual partner (an unknown but sympathetic adult for children), in second situation negatively and will display rejection towards a partner (hostile for children). The dyadic interaction between two actors will occur but only target stimuli will be captured. The final sequence (avatar with nonverbal behavioural displays) will then be presented from first person perspective (point of view of person interacting with target stimuli) to evoke the feeling of interaction (from point of view of the interacting partner - unique to our study). This will allow for the creation of virtual versions of the video stimuli using 3D avatars as target objects that have several advantages:

1. Avatars of average faces and bodies can replace the actors,

2. Physical traits of objects can be manipulated (we will create set of paedophilic target stimuli in paraphilia-related preferred age groups (based on current scientific research), possessing the typical physical features of the child in the appropriate age group and having age-typical friendly non-verbal behaviour displays),

3. Avatars can replace actors in phase of receptivity/copulation,

4. Avatars can be presented in 3D reality using oculus rift, evoking the feeling of real interaction.

To our knowledge, only one research team worldwide has already successfully employed avatars corresponding to different developmental stages in study of paraphilia.[14] For this study we will obtain three samples - volunteers with no paedophilic preference, volunteers with paedophilic preference but without history of child abuse (from paedophilic community we closely cooperate with) and patients with paedophilic disorder with history of child abuse who will voluntarily undergo the procedure. As part of the procedure we will measure physiological reactions such as penile tumescence, simultaneously with galvanic skin response, electro cardio graphy, and fMRI[15] data will be collected during the procedure to explore reactions in different participating groups.

## 3. Ethical Constrains

We are aware of numerous ethical controversies connected with the proposed research topics - especially because the study involves children-actors as participants and stimuli. We already went through large amount of consultations of stimuli content with forensic experts and judge advocates in field of sexology. We do have a permission of ethical committee[16] to run the study and it is supported by representatives of mental hospitals in field of sexology we closely cooperate with since there is urgent need for development of standardized set of stimuli for more accurate testing of paraphilic preferences.

In first study, we believe fewer controversies will occur since the aim of the study is generally non-conflicting with Czech law.

For study 2 child-actor- based avatars will not be presented in the erotic pose, the main focus is on child - typical nonverbal displays in avatars. Also we guarantee strict anonymization of data, in all parts of study and special attention will be focused on blindness of researches to individuals diagnoses, volunteer participation of all individuals, training of researchers/clinicians in techniques of communication on sexual topics/debriefing, intimacy assurance during testing and security arrangement during testing and after testing - skilled psychologist and psychiatric doctor will be present in all times and the supervising MD will be available for case of complications of any kind. Participants will be debriefed with non-interesting neutral film for purpose of lowering excitation of the participant lasting 30 minutes and will be contacted in time periods of 12 hours, 36 hours and one week after participation. No person will be forced to participate and everyone is allowed to end the procedure at any time. To avoid risk of unexpected complications during commuting, patients (and volunteers) will be transported to Institute on our expenses.

## Notes

[1] Matthew Price and Page Anderson, 'The Role of Presence in Virtual Reality Exposure Therapy', *Journal of Anxiety Disorders* 21.5 (2007): 742-751.

[2] In our case Kinect 2 - body, and Intel RealSense - facial expressions.

[3] Monica M. Moore, 'Nonverbal Courtship Patterns in Women: Context and Consequences', *Ethology and Sociobiology* 6.4 (1985): 237-247.

[4] Ibid, 237-247.

Karl Grammer, 'Strangers meet: Laughter and Nonverbal Signs of Interest in Opposite-Sex Encounters', *Journal of Nonverbal Behavior* 14.4 (1990): 209-236.

[5] Moore, 'Nonverbal Courtship Patterns in Women', 237-247.

Grammer, 'Strangers meet', 209-236.

[6] Erik Erikson, 'Eight Ages of Man', in *Childhood: Critical Concepts in Sociology*, (2005): 313.

[7] Age categories under 15; parents will obtain the detailed script of the scene before they underwrite the informed content about the participation of their child in the project

[8] Just introducing themselves in a way to gain the attention of the adult.

[9] Kurt Freund and Ray Blanchard, ‘Phallometric Diagnosis of Pedophilia’, *Journal of Consulting and Clinical Psychology* 57.1 (1989): 100.

[10] Adult man and woman, children of both sexes aged 6-7 years.

[11] Penile plethysmography (PPG): measurement of penile responses to stimuli that vary on the dimensions of interest. It is considered to be a specific measure of sexual arousal. For simultaneous measurement of PPG and fMRI we will use MR compatible custom-built penile plethysmography recording system HEINRICH SOM-04.

[12] Galvanic skin response (GSR): measuring electrical conductance of the skin using electrodes placed on a patient's body. Increase of sweat glands is considered to be signal of emotional excitement. We will use standard GSR equipment of Biopac systems.

[13] Electrocardiography (ECG): recording the electrical activity of the heart over a period of time using electrodes placed on a patient's body. Increase of electrical activity is considered to be signal of emotional excitement. We will use standard ECG equipment of Biopac systems.

[14] Renaud Patrice, Trottier Dominique, Rouleau Joanne-Lucine, Goyette Mathieu, Saumur Chantal, Boukhalfi Tarik and Stéphane Bouchard, ‘Using Immersive Virtual Reality and Anatomically Correct Computer-Generated Characters in the Forensic Assessment of Deviant Sexual Preferences’, *Virtual Reality*, 18.1 (2014): 37-47.

[15] Functional Magnetic Resonance Imaging (fMRI) is well-suited method for studying local and network brain functional correlates of psychological processes. fMRI uses Blood-Oxygen Level Dependent (BOLD) contrast, reflecting amount of oxygenated blood in an area, as an evidence of neural activity Measurements will be conducted using 3T Siemens device.

[16] On National Institute of Mental Health, Klecany, Czech Republic.

## Bibliography

Erik Erikson, ‘Eight Ages of Man’. In *Childhood: Critical Concepts in Sociology* (2005): 313.

Freund, Kurt, and Ray Blanchard. ‘Phallometric Diagnosis of Pedophilia’. *Journal of Consulting and Clinical Psychology* 57.1 (1989): 100.

Grammer, Karl. ‘Strangers meet: Laughter and Nonverbal Signs of Interest in Opposite-Sex Encounters’. *Journal of Nonverbal Behavior* 14.4 (1990): 209-236.

Moore, Monica M. ‘Nonverbal Courtship Patterns in Women: Context and Consequences’. *Ethology and Sociobiology* 6.4 (1985): 237-247.

Price, Matthew, and Page Anderson. ‘The Role of Presence in Virtual Reality Exposure Therapy’. *Journal of Anxiety Disorders* 21.5 (2007): 742-751.

Renaud Patrice, Trottier Dominique, Rouleau Joanne-Lucine, Goyette Mathieu, Saumur Chantal, Boukhalfi Tarik, and Stéphane Bouchard. ‘Using Immersive Virtual Reality and Anatomically Correct Computer-Generated Characters in the Forensic Assessment of Deviant Sexual Preferences’. *Virtual Reality* 18.1 (2014): 37-47.

**Binter, Jakub, M.A**., Ph.D. candidate at the Faculty of Humanities, Charles University in Prague, Czech Republic and researcher at the National Institute of Mental Health, Czech Republic. His research includes issues on sexual cognitions, situational hormonal response, vocal modulation, minor sexual preferences, and use of virtual reality in psychological research. Holder of Owen Aldis Scholarship 2014.

**Kateřina Klapilová, Ph.D.,** assistant professor at the Faculty of Humanities, Charles University in Prague, Czech Republic and researcher at the National Institute of Mental Health, Czech Republic. Her research includes issues on female sexuality, psychophysiological response, sexual cognitions, ethiology of paraphilia, and use of virtual reality in psychological research.

**Tereza Zikánová, M.A**., Ph.D. candidate at the Faculty of Humanities, Charles University in Prague, Czech Republic and researcher at the National Institute of Mental Health, Czech Republic. Her research includes issues on courtship non-verbal behaviour, situational hormonal response.

**Tommy Nilsson, MSc.**, researcher at the National Institute of Mental Health, Czech Republic. His research includes issues on Human Computer Interaction , virtual and mixed reality.

**Klára Bártová, M.A**., assistant professor at the Faculty of Humanities, Charles University in Prague, Czech Republic and researcher at the National Institute of Mental Health, Czech Republic. Her research includes issues on sexual orientation, evolutionary theory of human sexuality, partner choice, and use of virtual reality in psychological research.

**Lucie Krejčová, M.A**., Ph.D. candidate at the Faculty of Humanities, Charles University in Prague, Czech Republic and researcher at the National Institute of Mental Health, Czech Republic. Her research includes issues on female sexuality and psychophysiological measures of female sexual response.

**Renata Androvičová, M.A**., Ph.D. candidate at the Faculty of Humanities, Charles University in Prague, Czech Republic and researcher at the National Institute of Mental Health, Czech Republic. Her research includes issues on ethiology of paraphilia, and role of endocannabinoid system in regulation of sexual response.

**Jitka Lindová, Ph.D.,** assistant professor at the Faculty of Humanities, Charles University in Prague, Czech Republic and researcher at the National Institute of Mental Health, Czech Republic. Her research includes issues on non-verbal behaviour, especially synchrony and vocalization, interspecies communication, and cognitive psychology.

**Denisa Průšová,** Ph.D. candidate at the Faculty of Humanities, Charles University in Prague, Czech Republic and researcher at the National Institute of Mental Health, Czech Republic. Her research includes issues on behavioural patterns in romantic relationships in the context of power and dominance from a psychoterapeutical and communicational perspectives.

**Timothy Wells, Ph.D.,** assistant professor at the Faculty of Humanities, Charles University in Prague, Czech Republic and researcher at the National Institute of Mental Health, Czech Republic. His research includes multimodal communication and the perception of human attractiveness, with specific interest in quantitative research design and analysis.

**Riha, Daniel, Ph.D**., assistant professor at the Faculty of Humanities, Charles University in Prague, Czech Republic and researcher at the National Institute of Mental Health, Czech Republic. His research includes issues on serious games, virtual worlds and virtual reality. He is as well an award-winning artist (Kunst am Bau, Germany).

**Acknowledgement**

The author(s) disclosed receipt of the following financial support for the research, authorship, and/or publication of this chapter: Preparation of this manuscript was supported by the project "*National Institute of Mental Health (NIMH-CZ)*", grant number CZ.1.05/2.1.00/03.0078 (and the European Regional Development Fund), by the The Ministry of Education, Youth and Sports - Institutional Support for Longterm Development of Research Organizations - Charles University, Faculty of Humanities (Charles Univ, Fac Human 2014), by the grant SVV: 2014 - 260119 and SVV: 2015 - 260239.